\documentclass[12pt,preprint]{aastex}
\begin{document}

\parindent = 1.0cm

\title{A Large Group of AGB Stars in the Disk of M31: A Missing Piece of the Puzzle?}

\author{T. J. Davidge}

\affil{Herzberg Institute of Astrophysics,
\\National Research Council of Canada, 5071 West Saanich Road,
\\Victoria, BC Canada V9E 2E7}

\begin{abstract}
	The properties of a stellar grouping that is $\sim 3.5$kpc to the north 
east of the center of M31 is examined. This structure has (1) a surface 
brightness that is lower than the surrounding disk, (2) a more-or-less round 
appearance, (3) a size of $\sim 300$ arcsec ($\sim 1$ kpc), and (4) an integrated 
brightness M$_K = 6.5$. It is populated by stars with ages $\geq 100$ Myr and 
$J-K$ colors that tend to be bluer than those of stars in the surrounding disk. 
Comparisons with model luminosity functions suggest that the star formation rate 
in this object has changed twice in the past few hundred Myr. Fitting a Sersic 
function to the light profile reveals a power-law index and effective surface 
brightness that are similar to those of dwarf galaxies with the same integrated 
brightness. Two possible origins for this object are considered: (1) it is 
a heretofore undiscovered satellite of M31 that is seen against/in/through the 
M31 disk, or (2) it is a fossil star-forming region in the M31 disk.
\end{abstract}

\keywords{galaxies: individual (M31) --- galaxies: evolution --- galaxies: spiral}

\section{INTRODUCTION}

	The tidal streams that lace the extraplanar regions of M31 (Ibata et al. 
2007; Tanaka et al. 2010) are testament to past interactions with neighboring 
systems. The tidal debris and companion galaxies that surround M31
are pieces of a puzzle that -- when all the bits 
of information are assembled -- should form a coherent picture of the evolution 
of the galaxy. A key goal is to determine if the evolution of M31 has been 
driven by a single major merger, as suggested by Hammer et al. (2010), or by 
the repeated pummelling by smaller satellites.

	The disk of M31 contains additional pieces of the M31 evolution puzzle. 
The kinematic properties (Unwin 1983) and spatial distribution 
(Yin et al. 2009; Nieten et al. 2006) of star-forming material in M31 is 
indicative of a disturbed gas disk, and star formation at the present day 
occurs mainly in arcs and ring-shaped structures 
(Gordon et al. 2006; Barmby et al. 2006), rather than large-scale spiral arms. 
The age distribution of globular clusters is suggestive of 
at least two significant star-forming events in the disk since the 
initial assembly of the galaxy (Puzia et al. 2005). The M31 disk also has a 
significantly larger spatial extent than the Galactic disk (Hammer et 
al. 2007), as expected if the M31 disk has accreted material and been `spun up'. 

	An investigation of the central few kpc of the M31 disk is of 
interest as it might be an area of past and/or present star-forming 
activity given that large-scale interactions can channel gas into 
the inner regions of galaxies. Dynamical friction 
might also drag satellites into the central regions of galaxies; indeed, 
Saglia et al. (2010) find tantalizing hints that a satellite was accreted by M31 
$\sim 100$ Myr ago. In the current paper, we report on the detection of a prominent 
structure in the inner disk of M31.

\section{DATA}

\subsection{2MASS images}

	$J$ and $Ks$ images of M31 from the 2MASS Large Galaxy Atlas 
(Jarrett et al. 2003) are used to search for localized structures that depart from 
the large-scale isophotal properties of the disk. The goal of this exercise is not 
to characterize features such as bars or asymmetries in the disk 
that have spatial scales approaching a disk scale 
length; rather, the search focuses on structures with characteristic sizes 
$\leq 1 - 2$ kpc. The 2MASS images trace the light profile of M31 out to 30 arcmin 
(R$_{GC} \sim 6$ kpc), and so the coverage is restricted to the inner disk of 
the galaxy.

\subsection{WIRCam Images}

	The photometric properties of stars in a single $20 \times 20$ arcmin$^2$ 
CFHT WIRCam (Puget et al. 2004) field, which samples the major axis to the north 
east of the galaxy center, were examined to characterize an object that was 
discovered in the 2MASS images. The data consist of forty 20 second $J$ and 
$Ks$ images recorded for joint programs 2009BC29 and 2009BH52. Individual 
background-subtracted images processed with the I`IWI pipeline were downloaded 
from the CADC CFHT archive, and these were spatially 
registered to correct for exposure-to-exposure jitter prior to stacking. 
Stars in the final combined images have FWHM = 0.7 arcsec.

\section{RESULTS}

\subsection{The Identification of a Large Structure}

	The isophote-fitting program $ellipse$ (Jedrzejewski 1987) was run on 
the 2MASS images after they had been smoothed with a running boxcar median 
filter to suppress bright stars. A model constructed from the fitted isophotes 
was subtracted from the images, and stellar concentrations with sizes $\leq$ a few 
hundred arcsec will appear as excess light in the result. The left hand panel of 
Figure 1 shows the isophote-subtracted LGA $J$ image. 
A number of faint residual structures are evident 
throughout the $85 \times 85$ arcmin$^2$ field, including the central cross-shaped 
pattern that is a signature of the boxy nature of the M31 bulge (Beaton 
et al. 2007). Aside from NGC 205 and M32, by far the most pronounced diffuse 
structure in the residual image is $\sim 17$ arcmin ($\sim 3.5$ kpc) to 
the north east of the galaxy center. This object is centered at RA = 00:43:43 and 
Dec = 41:28:15 (E2000). It is outside the region where light from the bulge 
dominates at visible wavelengths (e.g. Tempel et al. 2011), and is near the end 
of the bar identified by Beaton et al. (2007). A $10 \times 10$ arcmin$^2$ section 
of the isophote-subtracted LGA $J$ image centered on this object is shown in the 
lower right panel, and it can be seen that the central regions have a more-or-less 
round morpholgy. In subsequent discussion we refer to this feature as the NE Clump.

	While the surface brightness of the NE Clump is 
$\sim 1.3$ mag arcsec$^{-2}$ fainter than the surrounding disk, it is 
still a significant structure, with integrated brightnesses 
$J = 7.3$ and $K = 6.5$. For comparison, the integrated apparent brightness 
of M32 is $K = 5$, while for NGC 205 $K = 5.5$ (Jarrett et al. 
2003); the NE clump is thus a large -- heretofore unidentified -- 
structure in the M31 system. Adopting a distance modulus of 24.36 
(Vilardell et al. 2010) then M$_K \sim -17.9$.

	The overall appearance of the NE Clump is suggestive of a low surface 
brightness galaxy. The surface brightness and color profiles of the NE clump 
are shown in Figure 2. As demonstrated in the upper panel of Figure 2, the 
$J$ light profile of the NE Clump can be well matched by a Sersic profile 
with an exponent $n = 0.48 \pm 0.03$, R$_e = 169 \pm 4$ arcsec 
(i.e. $593 \pm 14$ parsecs), and $\mu_e = 20.37 \pm 0.02$ mag arcsec$^{-2}$. 
The $J-K$ color in the central 200 arcsec is $\sim 0.4$ magnitudes bluer than 
that of the surrounding disk, and is similar to that of NGC 185, which has a 
similar $K-$band brightness (Jarrett et al. 2003).

	Some of the structural properties of the NE Clump are not similar to those 
of nearby dwarf galaxies. We have compared the Sersic parameters of the NE Clump 
with those compiled by Jerjen et al. (2000) for nearby dwarf galaxies. 
To make these comparisons we note that the $J-K$ color of the NE Clump 
is suggestive of $B-K \sim 3 - 4$, and so M$_B \sim -14$ to $-15$. Comparing the 
Sersic parameters of the NE Clump with the relations shown in Figures 5 and 6 
of Jerjen et al. (2000): (1) $n$ is near the low end of dwarf galaxy 
values in this M$_B$ range, but still within the scatter envelope, (2) 
the central surface brightness is well within the envelope defined by nearby 
dwarfs, but (3) R$_e$ is $\sim 0.3$ dex larger than for other dwarfs.

\subsection{The Stellar Content of the NE Clump}

	The WIRCam data samples the southern half of the NE Clump. 
Stellar brightnesses in the WIRCam image were measured with the PSF-fitting 
routine ALLSTAR (Stetson \& Harris 1988), using PSFs and source catalogues 
constructed with routines in the DAOPHOT (Stetson 1987) 
package. The $(K, J-K)$ CMD of stars in the 
NE Clump area indicated in Figure 1 is shown in the lower panel 
of Figure 3. The peak brightnesses of the red supergiant (RSG) and 
asymptotic giant branch (AGB) sequences in the Z = 0.019 Marigo et al. 
(2008) isochrones are indicated at the top of the figure. The brightest 
AGB stars in these models have $K \sim 14.2$. There is a smattering of stars 
brighter than this, and these may be long period variables near the peak 
of their light curves and/or RSGs with ages log(t) $< 7.3$.

	Contamination from the M31 disk complicates efforts to probe the stellar 
content of the NE Clump. Here, statistical corrections for disk contamination are 
based on number counts in control fields that are near the NE clump, and the 
locations of these fields are indicated in the upper 
right hand panel of Figure 1. The $K$ luminosity functions 
(LFs) of objects with $J-K$ colors between 0.6 and 1.8 were constructed for the NE 
Clump and the control fields. The control field LFs were combined and scaled to 
match the disk surface brightness near the NE Clump using the isophotal 
measurements described in Section 3.1. These scaled counts 
were then subtracted from the observed NE Clump LF to produce the LF shown in the 
middle panel of Figure 3; the observed (i.e. NE Clump $+$ 
disk component) LF is shown in the top panel of Figure 3.

	Stars remain in the NE Clump area after disk subtraction, indicating 
that the NE Clump stands out as a distinct feature in 
star counts. Given that (1) the mean surface 
brightness of the NE Clump is over a magnitude arcsec$^{-2}$ lower than the 
surrounding disk (\S 3.1) and (2) disk subtraction removes roughly one half 
of the stars, then it appears that the NE Clump contains a 
higher fraction of bright AGB stars per unit near-infrared surface brightness than 
the surrounding disk. This could indicate a higher specific star 
formation rate (SFR) in the NE Clump when compared 
with the disk during the past few hundred Myr, in agreement with the 
difference in the color distributions of the NE Clump and the disk (see below).

	Model $K$ LFs for simple stellar populations 
with Z = 0.019 were constructed from the Marigo et al. (2008) isochrones using the 
tools provided on the {\it Padova database of evolutionary tracks and isochrones} 
web site \footnote[3]{http://stev.oapd.inaf.it/cgi-bin/cmd}, and these were 
combined to construct a $K$ LF for a system with a constant SFR in the time 
interval log(t$_{yr}$) = 8.1 to 8.5. The Kroupa et al. (1993) 
mass function was adopted. The resulting model, normalized to match the 
observations with $K$ between 14.5 and 15.0, is shown as a dashed line in the 
middle panel of Figure 3. The model does not match the observations. 
The departures between the model and observations suggest that the 
SFR of the NE Clump changed at log(t$_{yr}$) = 8.15 and again at 
log(t$_{yr}$) = 8.35.

	The color distribution of AGB stars provides additional 
information about the stellar content of the NE Clump. The $J-K$ distribution 
of stars with $K$ between 14.5 and 15.0 in the NE Clump is examined in Figure 4. 
The raw color distribution (i.e. with no correction for disk contamination) 
is shown in the top panel, while the distribution after 
subtracting the contribution from the control fields is shown in the lower panel. 
Monte Carlo simulations, in which the number 
counts in each color bin of the NE Clump and control fields were perturbed by 
the uncertainties predicted from counting statistics, 
were run to determine the significance of the differences between the 
color distribution of the NE Clump and the control fields. These 
simulations indicate that the peak of the $J-K$ distribution in the NE Clump 
is bluer than that in the control fields at more than the 2--$\sigma$ 
significance level. The color distribution of AGB stars in the NE Clump is thus 
skewed to bluer colors than the disk contribution, indicating that 
the stellar mixture in the NE Clump differs from that in the surrounding disk. 
The required differences in stellar content can be estimated from 
the Marigo et al. (2008) models, which indicate that a change of 0.05 magnitudes in 
the peak of the J--K color distribution corresponds to $\Delta$[M/H] $\sim 0.3$ 
dex or a difference in effective age $\Delta$log(t$_{yr}$) = 0.5 dex.
 
\section{DISCUSSION}

	A concentration of AGB stars has been identified on the major axis 
of M31 3.5 kpc to the north and east of the galaxy center. The `NE Clump' is 
evident in both integrated near-infrared light and star counts. The NE Clump has 
probably eluded detection to date because much 
of the previous work on M31 has focused on either the 
central areas of the bulge or the outermost regions of the galaxy.
In addition, structures of this nature are difficult to detect at visible 
wavelengths given their low surface brightness and sub-structure introduced by 
blue stars and dust. However, at near-infrared wavelengths it is possible to 
probe the stellar content to comparatively small radii due 
to (1) the reduced levels of dust extinction, and (2) the diminished effects 
of line blanketing at these wavelengths, which results in greater contrast 
between luminous red giant branch stars and the underlying body of bluer stars. 

	The NE clump is dominated by stars with ages $\geq 100$ Myr, and is not a 
simple stellar system. Indeed, comparisons with a model LF suggest that the SFR 
changed 140 Myr and 220 Myr in the past. The $J-K$ color distribution of AGB stars 
in the NE Clump is skewed to blue colors when compared with the distribution of the 
surrounding disk, suggesting that stars in the NE Clump have a lower metallicity 
and/or a younger mean age than those in the surrounding M31 disk.

	We consider two possible origins for the NE Clump:

\vspace{0.3cm}
\noindent{1)} The NE Clump is a heretofore unidentified satellite of M31. 
The Sersic index and central surface brightness of the NE Clump fall within the 
range defined by nearby dwarf galaxies. In contrast, the effective radius of 
the NE Clump is larger than expected for a dwarf galaxy. Depending on the 
distance above/below the M31 disk plane, the size and 
structure of the NE Clump may be affected by tidal forces. If the NE Clump is a 
dwarf galaxy that contains stars with a wide range of ages then M/L$_K \sim 1$, and 
its stellar mass will be $\sim 3 \times 10^8$ M$_{\odot}$. Given that the mass of 
M31 within 5 kpc of the nucleus is $\sim 6.7 \times 
10^{10}$ M$_{\odot}$ (e.g. Carignan et al. 2006), 
then the tidal radius of the NE Clump is $\sim 330$ parcsecs ($\sim 95$ arcsec) 
if it is in the M31 disk plane. The presence of dark matter in the NE Clump 
would push the tidal radius to higher values.
Finally, while the colors of AGB stars in the NE Clump and its 
comparatively blue integrated $J-K$ color are consistent with 
a stellar mixture that differs from that of the M31 disk, it is not clear if 
this is due to differences in metallicity, age, or both. A difference 
in metallicity would be expected if the NE Clump is a dwarf galaxy. 

\vspace{0.3cm}
\noindent{2)} The NE Clump is a fossil star-forming region in the M31 disk. 
The NE Clump has experienced star formation for an extended period of time, 
and there is evidence of changes in the SFR during the past $\sim 200$ Myr 
ago. There are other areas of recent star formation in the disk of M31 that have 
been active for a similar extended period of time (e.g. Davidge et al. 2012), 
and Williams (2003) and Davidge et al. (2012) conclude that the 
area $5 - 10$ kpc to the north east of the center of M31 
had elevated levels of star formation during the past few hundred Myr. One possible 
trigger for star formation over kpc spatial scales is an interaction 
with a companion. In fact, the change in SFR in the NE Clump 220 Myr ago 
coincides with the timing of the passage of a satellite through the inner disk of 
M31 proposed by Block et al. (2006), as well as with 
the formation of the Giant Stellar Stream (GSS), which models suggest may 
have formed between 0.25 Gyr (Font et al. 2006) and 0.65 Gyr (Fardal et al. 
2007) in the past. In contrast, Hammer et al. (2010) suggest that the 
GSS formed much earlier, due to an interaction between M31 and another large 
galaxy.

	The NE Clump is near the end of the bar identified by Beaton 
et al. (2007), and the compression of gas as the bar passes through the disk may 
induce star formation. However, the more-or-less round morphology of the 
NE Clump may not be consistent with such as trigger. Dobbs \& Pringle 
(2010) investigate the distribution of stars in a barred spiral galaxy, and 
stars in their models that formed in the bar $\sim 50 - 100$ Myr in the past are 
smeared azimuthally, with additional radial blurring.

	There are two observational predictions that 
could be tested to probe the nature of the NE Clump. 
If it is a dwarf galaxy then (1) spectroscopic studies should 
find that a substantial fraction of stars in this field are 
chemically distinct from those of the disk, and (2) deep high angular resolution 
imaging should reveal an RGB sequence that is skewed to bluer colors than in the 
surrounding disk. Spectroscopic studies may also find that any stars that are 
chemically distinct from those in the disk also have non-disk kinematics, 
although this depends on the orbit of the NE Clump.

	If it is found to be a dwarf galaxy then the NE Clump is a new 
consideration when interpreting structures in the M31 system. When its location and 
properties with respect to the M31 disk are established then it may prove to be the 
source of at least some of the tidal features that are seen today. Indeed, some 
structures, such as the GSS and the kinematically distinct 
nuclear gas ring (Saglia et al. 2010), do not have an identified progenitor.

\acknowledgements{Thanks are extended to the anonymous referee, who provided 
comments that greatly improved the paper.}

\clearpage

\begin{figure}
\figurenum{1}
\epsscale{1.0}
\plotone{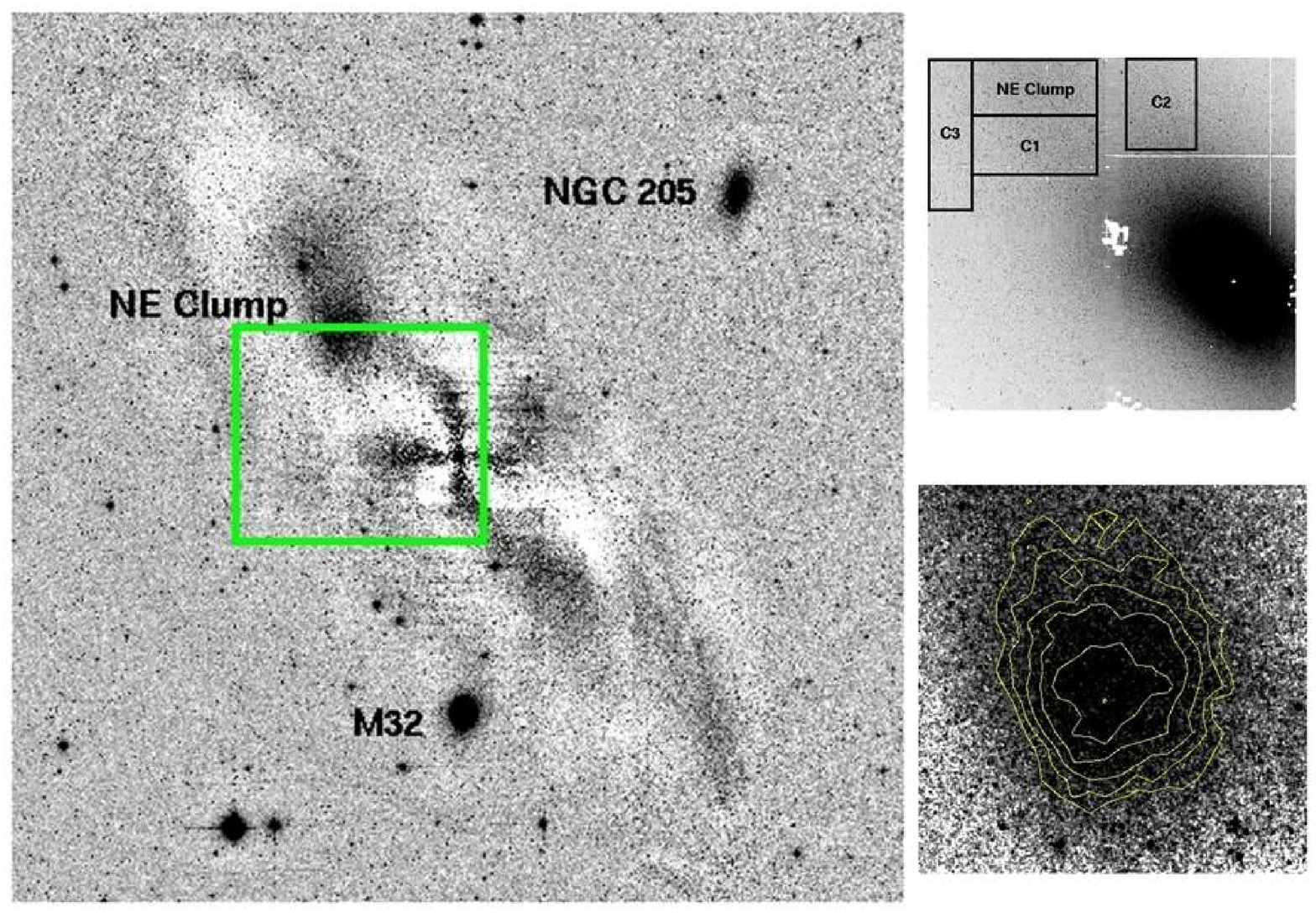}
\caption{(Left panel) The residual image produced by 
subtracting an isophotal model from the 2MASS $J$ LGA 
image of M31. North is at the top, and East is to the left. An 
$85 \times 85$ arcmin$^2$ area is shown, with the nucleus of M31 at the center. 
The central cross-shaped pattern is an artifact of the boxy 
structure of the M31 bulge (Beaton et al. 2007). The NE Clump is the concentration 
of light $\sim 17$ arcmin ($\sim 3.5$ kpc) to the north east of the galaxy 
center. The location of the WIRCam field that is used to probe the stellar content 
of this structure is indicated with the green box. (Lower right panel) The 
$10 \times 10$ arcmin$^2$ portion of the isophote-subtracted LGA $J$ image centered 
on the NE Clump. The outermost contour is $\mu_J = 20.9$ mag arcsec$^{-2}$, 
and the contour interval is 0.3 mag arcsec$^{-2}$. (Upper right panel) The final 
$K$ WIRCam image, which covers $20 \times 20$ arcmin$^2$ and samples the 
southern half of the NE Clump. The areas used in the photometric analysis to 
probe the NE Clump and the surrounding disk are indicated.}
\end{figure}

\clearpage
\begin{figure}
\figurenum{2}
\epsscale{1.0}
\plotone{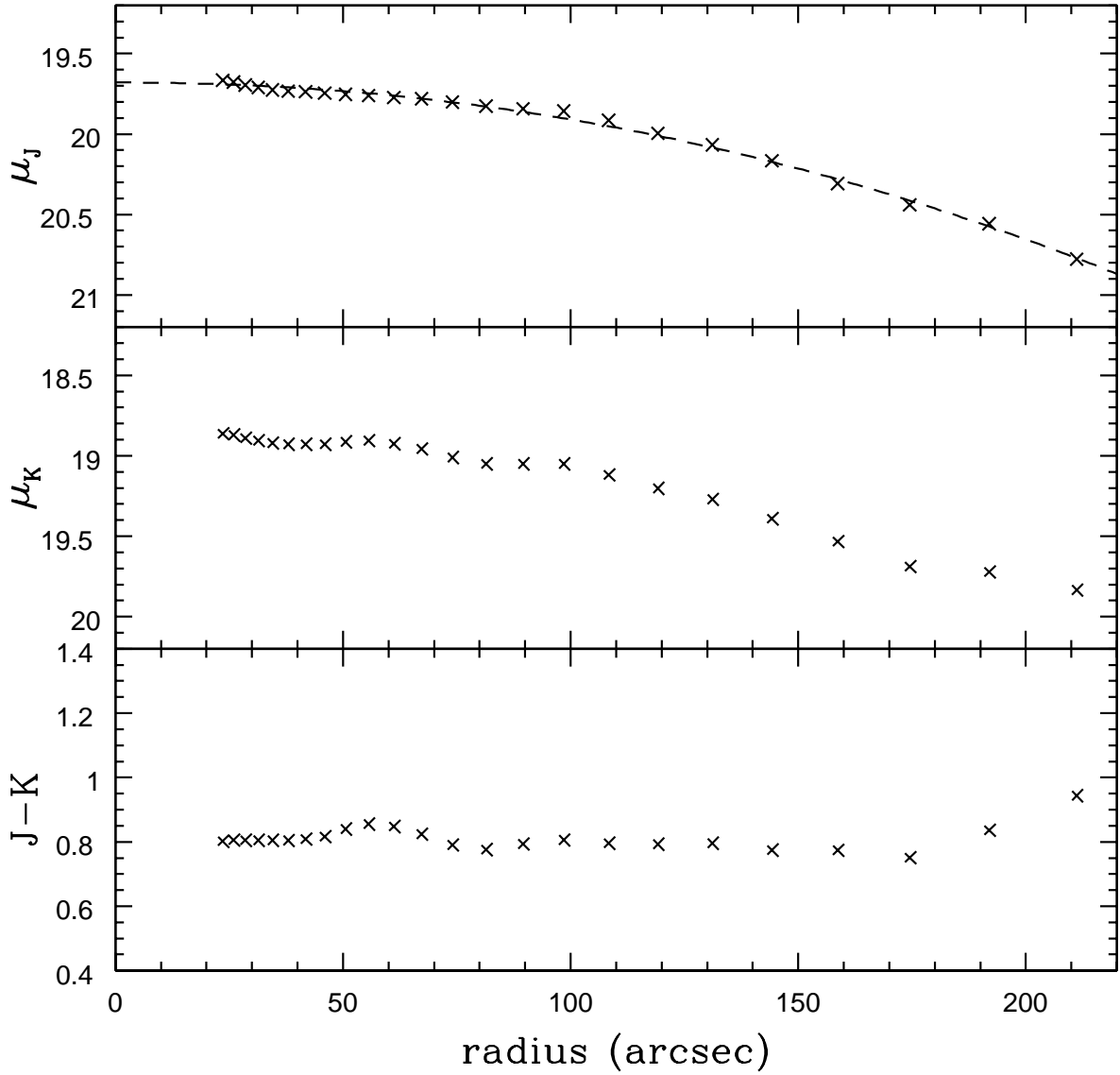}
\caption{The surface brightness and color profiles of the NE Clump, measured from 
the isophote-subtracted 2MASS LGA images. Surface brightnesses are in mag 
arcsec$^{-2}$. The dashed line in the top panel shows a Sersic profile with n = 
0.48, r$_e = 169$ arcsec, and $\mu_e = 20.366$ mag arcsec$^{-2}$.}
\end{figure}

\clearpage
\begin{figure}
\figurenum{3}
\epsscale{1.0}
\plotone{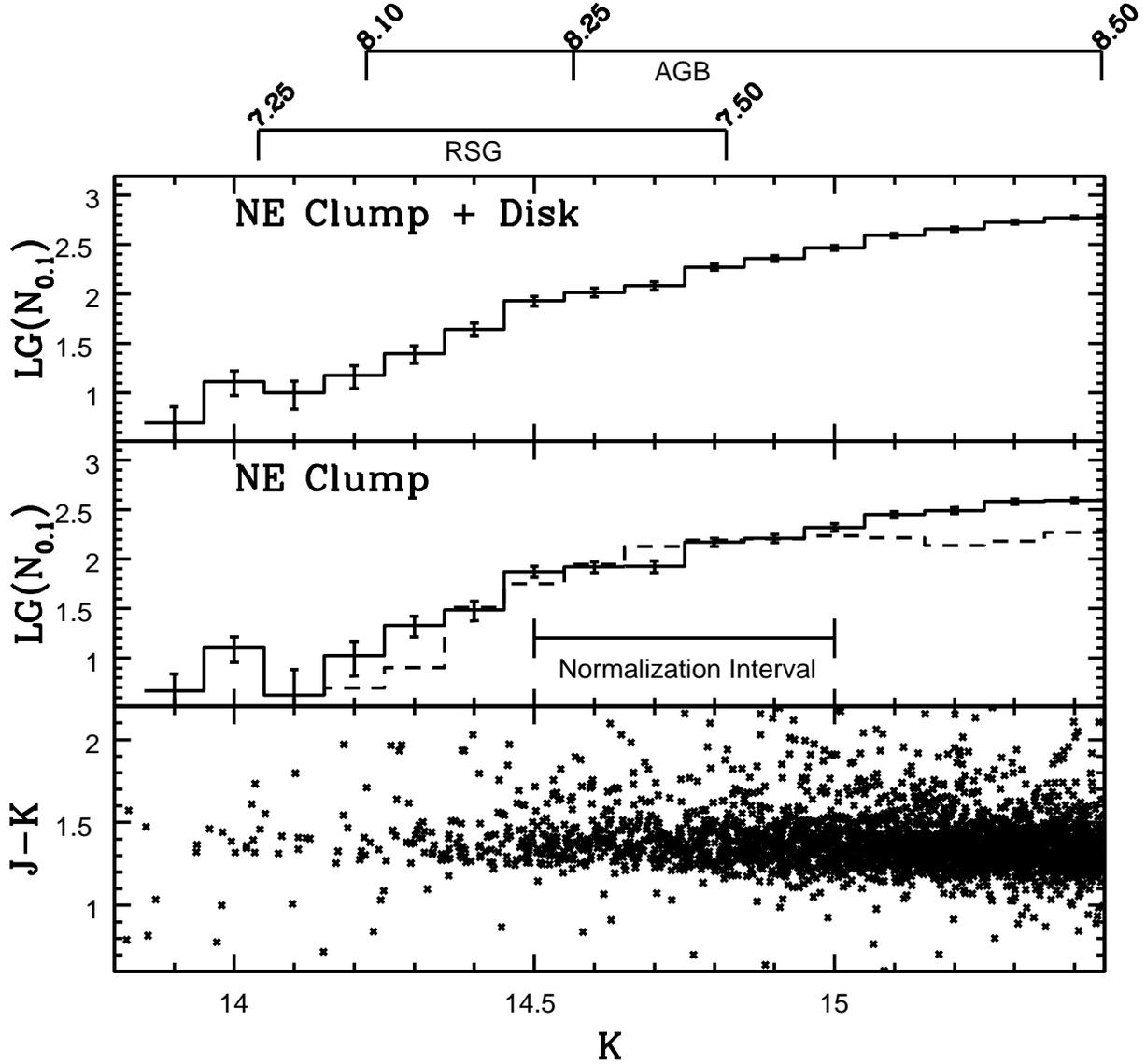}
\caption{The photometric properties of objects in the NE Clump. 
The peak brightnesses of RSGs and AGB stars predicted from the Marigo et al. 
(2008) isochrones are shown at the top of the figure, with log(t$_{yr}$) 
indicated. (Top panel) The observed $K$ luminosity 
function (LF) of the NE Clump, including contamination from the M31 disk. 
N$_{0.1}$ is the number of stars with $J-K$ between 0.6 and 1.8 per 0.1 magnitude 
interval. (Middle panel) The $K$ LF of the NE Clump after 
subtracting star counts from control fields in the disk. 
The dashed line is a model LF constructed from the 
Marigo et al. (2008) Z = 0.019 isochrones assuming a constant SFR and a 
Kroupa et al. (1993) mass function -- the model 
has been normalized to match the observations between $K = 14.5$ and 15. There 
is poor agreement with the observations, and the SFR in the NE Clump appears to 
have changed near log(t$_{yr}$) = 8.15 and again at log(t$_{yr}$) = 8.35. 
(Bottom Panel) The CMD of the NE Clump, including contamination from the disk. 
A well-populated AGB is present with $J-K \sim 1.4$.} 
\end{figure}

\clearpage
\begin{figure}
\figurenum{4}
\epsscale{1.0}
\plotone{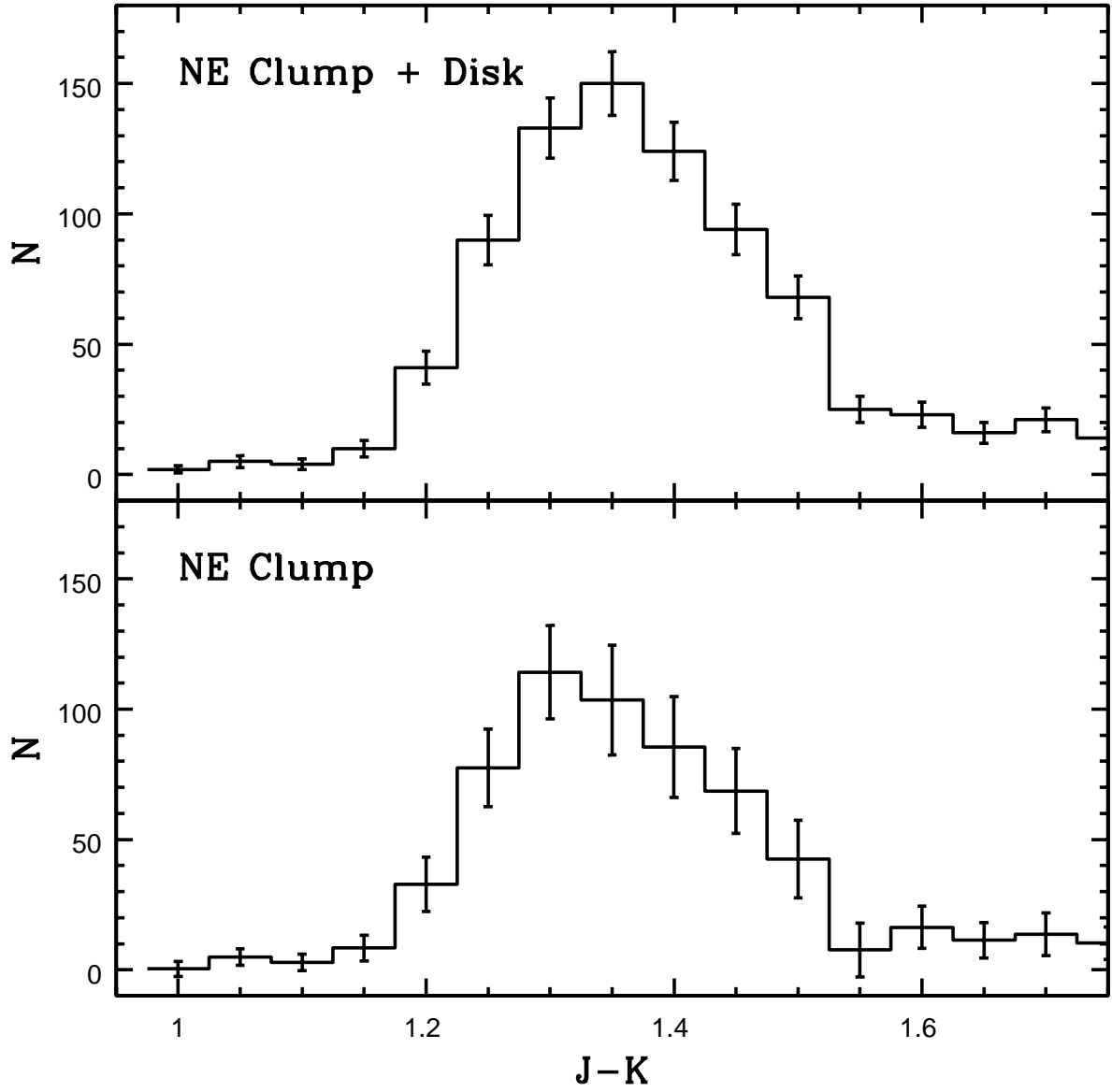}
\caption{(Top panel) The distribution of the $J-K$ colors of stars with $K$ 
between 14.5 and 15.0 in the NE Clump, including disk contamination. 
(Lower Panel) The color distribution after correcting for 
disk contamination. The color distribution in the lower panel is 
skewed to bluer values than that of the control fields, indicating that 
the stellar mixture in the NE Clump differs from that in the surrounding M31 disk.}
\end{figure}

\end{document}